\documentclass[letterpaper, 10 pt, conference]{ieeeconf}  

\IEEEoverridecommandlockouts                              

\usepackage{graphicx}
\title{\LARGE \bf
Big Data Spark Solution for Functional Magnetic Resonance Imaging 
}

\author{Saman Sarraf$^{ 1,2,3}$ and Mehdi Ostadhashem
\thanks{$^{1}$Saman Sarraf was with the Department of Electrical and Computer Engineering, McMaster University, Hamilton, ON, L8S 4L8, Canada 
		{\tt\small samansarraf@ieee.org}}%
\thanks{$^{2}$Saman Sarraf was with Rotman Research Institute at Baycrest, Toronto, ON, M6A 2E1, Canada}%
\thanks{$^{3}$Saman Sarraf is a full member of IEEE and EMBS.}
}

\begin{document}

\maketitle
\thispagestyle{empty}
\pagestyle{empty}

\begin{abstract}

Recently, Big Data applications have rapidly expanded into different industries. Healthcare is also one the industries willing to use big data platforms so that some big data analytics tools have been adopted in this field to some extent. Medical imaging which is a pillar in diagnostic healthcare deals with high volume of data collection and processing. A huge amount of 3D and 4D images are acquired in different forms and resolutions using a variety of medical imaging modalities. Preprocessing and analyzing imaging data is currently a long process and cost and time consuming. However, not many big data platforms have been provided or redesigned for medical imaging purposes because of some restrictions such as data format. In this paper, we designed, developed and successfully tested a new pipeline for medical imaging data (especially functional magnetic resonance imaging - fMRI) using Big Data Spark / PySpark platform on a single node which allows us to read and load imaging data, convert them to Resilient Distributed Datasets in order manipulate and perform in-memory data processing in parallel and convert final results to imaging format while the pipeline provides an option to store the results in other formats such as data frame. Using this new solution and pipeline, we repeated our previous works in which we extracted brain networks from fMRI data using template matching and sum of squared differences (SSD) method. The final results revealed our Spark (PySpark) based solution improved the performance (in terms of processing time) around 4 times on a single compared to the previous work developed in Python.  

\end{abstract}

\section{INTRODUCTION}

Imaging modalities produce significant amount of data. For example, in functional MRI (fMRI) which is one of most important of neuroimaging methods, we capture Blood-oxygen-level dependent (BOLD) signals of the whole brain across time. This approach collects three dimensional data (x,y,z) over time (t) and store four dimensional (4D) data \cite{sarraf2016functional}. Imaging data preprocessing and analysis is expensive in terms of providing infrastructure and processing time. A huge amount of data need to be analyzed which are producing a huge amount of results. Furthermore, the results need to be stored in disks and will be potentially retrieved for more data analyzes. One of the challenges in this filed is how the current medical image processing tools would migrate to use big data resources more efficiently as big data analytics platforms have been developed to ?massage? large datasets. Therefore, development of a pipeline enabling us to merge those two concepts (big data and imaging) seems to be necessary.  This potential pipeline allows us to read imaging data in an environment that can speak with a big data platform. Next, using features of the big data platform, the data manipulation such as data analyzes are performed in parallel which is not only improving the processing times, but it can also handle large datasets easily. Traditionally, results of analyzed medical images are stored in any standard medical imaging format. However, the developed pipeline provides the option of storing results in other formats such as data frame that can be easily read and written by other big data analytics tools.

\section{BACKGROUND AND ALGORITHMS}

\subsection{Big Data}

Big data is a developing term which is describing any voluminous amount of structured, semi-structured and unstructured data having the potential to be mined for information. To describe the Big Data phenomena four V\textquoteright s are used which are Volume, Velocity, Variety and Veracity \cite{sagiroglu2013big}\cite{mcafee2012big}. Volume refers to the huge amount of data generated every moment. All kind of data from Internet industry, sensor and machine data to healthcare are collected and stored. The traditional analytics methods are not able to handle this vase amount of data. In addition, storing and retrieving big data need more developed infrastructure. By using big data technology, we are be able to store and utilized these data sets with the help of distributed systems, where parts of the data is stored in different locations, connected by networks and brought together by software. Velocity in big data means the speed at which new data is generated and the speed of data retrieval and analysis on which we focus in the paper. Big data technology enable us to analyze the data in both offline and real-time while data stream is not even stored in database. Variety refers to the different types of data that are generated and used for different purposed such as big data analytics. Over past decade, not only the volume of data recorded has dramatically increased, but also the different types of data and new variety of data have been also collected.  In the past, the focused was more on structured data that fits into tables or relational databases such as financial or healthcare date. However, around 80 percent of the world?s data is now unstructured and therefore need big data tools to easily be stored into new style tables or databases. For example, how we can collect imaging data in research centers and store them in traditional databases? Big data technology has reliable solutions for differed types of data including sensor data or healthcare related data. This technology brings the unstructured or semi-structured data together with more traditional, structured data. Veracity speaks of trustworthiness of the data. The accuracy and precision of collected data affect the data analyses. There are a variety of forms of big data in which quality and accuracy are less controllable. Also, the volumes sometimes cause the lack of quality or accuracy. Big data tools and analytics technology now provide us to analyze these types of data Big Data \cite{sagiroglu2013big}\cite{mcafee2012big} .

\begin{figure}[h!]
	\includegraphics[width=\linewidth]{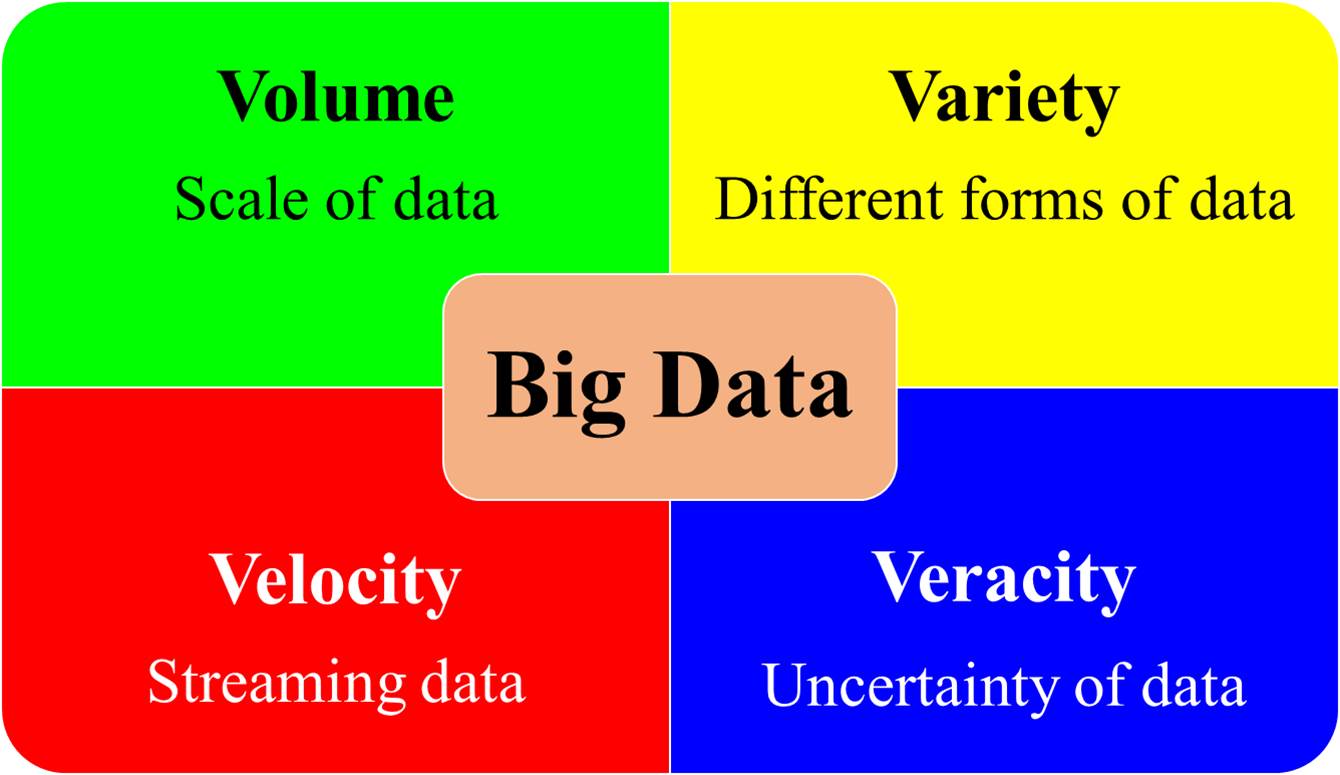}
	\caption{4 Vs in big data}
	\label{fig1}
\end{figure}

\subsection{Big Data in Healthcare}

In healthcare industry, a huge volume of data is generated in different ways. Electronic health records, imaging data, bio-signals are some of well-known types of data that are collected and stored in healthcare. Although the healthcare dare are usually stored as structured data, lack of golden standard of recording data in different healthcare divisions causes a variety of semi or unstructured data are stored. Big data analytics technology provide infrastructure to categorize and analyze the healthcare data \cite{viceconti2015big} \cite{andreu2015big}. Furthermore, by using this technology, novel and more accurate predictive models \cite{sarraf2014mathematical} are generated and complicated patterns are extracted from the big data \cite{andreu2015big}. One of the popular areas in healthcare that is dealing with all 4 Vs of big data is medical imaging and imaging informatics. Different imaging modalities, a variety of acquisition times and resolutions, and several imaging data formats and finally, different source of noise in data clearly state that medical imaging cover 4 Vs. However, there are few big data tools developed for medical imaging. With lack of cutting-edge technology, the imaging data are often processed and analyzed in classic way in which the performance (speed) is not high enough. Design of new processing and analysis pipeline using Big Data tools seems to be necessary these days.

\subsection{Big Data tools - Spark}

Big data technologies and tools such as Hadoop or Apache? Hadoop are open-source software programming platform and projects for reliable, scalable, distributed computing [hadoop.apache.org]. The Apache Hadoop software library is a framework installed on specific hardware infrastructure enables developers and users for the distributed and parallel processing of large data sets across clusters and nodes by using programming models. In theory, Hadoop is designed to function in single servers or thousands of nodes performing local computation and storage [hadoop.apache.org]. This platform included several sub-projects such as HDFS, MapReduce, YARN and etc. \cite{kala2013review}. However, some of the difficulties in using Hadoop framework such as complicated installation process and highly dependency on hardware structure were motivations to develop other Big Data platforms. Also, Hadoop supported few programming languages and it is not user friendly for data analysts not having programming background. Therefore, Big Data platform developers aimed to develop a software library supporting more programming languages, having less dependency on hardware and being more memory efficient. 
Spark and Apache Spark is one of the modern big data platforms [http://spark.apache.org/] which was developed originally at the University of California, Berkeley's AMPLab. Apache Spark? is practically a fast and general engine for big data processing. In memory data processing of Spark improved the performance up to 10 times faster than on-disk data processing. In addition, Spark offers over 80 high-level applications that can interactive with different programming languages such as Java, Scala, R and Python (which is more important in this paper). This feature shows the Spark ease of use compared to Hadoop platform. Another feature of Spark is ?Generality?. Spark can easily handle data streaming and real-time data processing. Also, it can easily interact with SQL databases and data frames. Machine learning library of Spark (MLlib) and GraphX which is the graph analysis tool of this plat form allow users to perform data processing and analysis in a fast and parallelized environment that can be installed either on single node as a standalone version or installed on thousands of nodes.

\subsection{Functional MRI - Brain Networks}

Functional Magnetic Resonance Imaging (fMRI) is a technique that measures brain activity by detecting the associated changes in blood flow. This MRI technique uses the change in magnetization between oxygen-rich and oxygen-poor blood in the brain as its primary outcome measure, with greater consumption of oxygen corresponding to greater neural recruitment within the brain \cite{sarraf2016functional} \cite{sarraf2014brain}. Our brain is an efficient network to be precise. It is a network made up of a large number of brain regions that have their own task and function but remain highly interactive by continuously sharing information with each other \cite{grady2016age} \cite{strother2014hierarchy}. As such, they form a complex integrative system in which information is continuously processed and transferred between structurally and functionally linked brain regions: the brain network \cite{sarraf2016functional} \cite{sarraf2014brain}.
\begin{figure}[h!]
	\includegraphics[width=\linewidth]{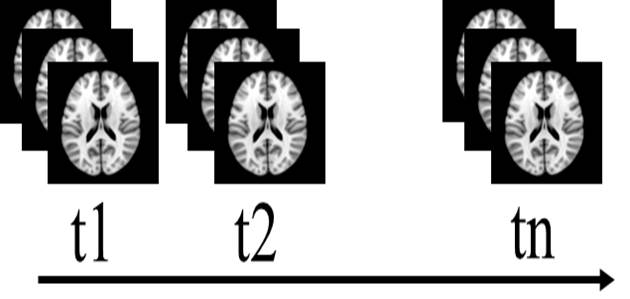}
	\caption{fMRI data acquisition 3D brain volume across time}
	\label{fig2}
\end{figure}
The data which are collected in fMRI modality are four dimensional (4-D) as volumes images are acquired across time. The preprocessing and analysis of this huge volume of data is always time consuming and costly required parallelized infrastructure. Therefore, the fMRI data analysis deals with at least 2 Vs of big data analytics: Volume and Velocity. In small to large fMRI datasets, over Giga to Tera bytes of data are preprocessed and analyzed. Also, every day, imaging research and healthcare centers collect more and more data and archiving this volume of data has become challenging. In addition, data format is another issue causing some restrictions to use big data platform in imaging. None of imaging data formats can be directly stored in database and be preprocessed or analyzed by standard big data analytics tools. To merge big data analytics and medical imaging, we proposed and developed a pipeline that can be used on single node PC or huge clusters and is able to process data much faster the current methods and also enables users to store the pre/post processed data in different data format compatible with big data plat forms especially Spark.
\section{RESULTS AND DISCUSSION}
\begin{figure}[h!]
	\includegraphics[width=\linewidth]{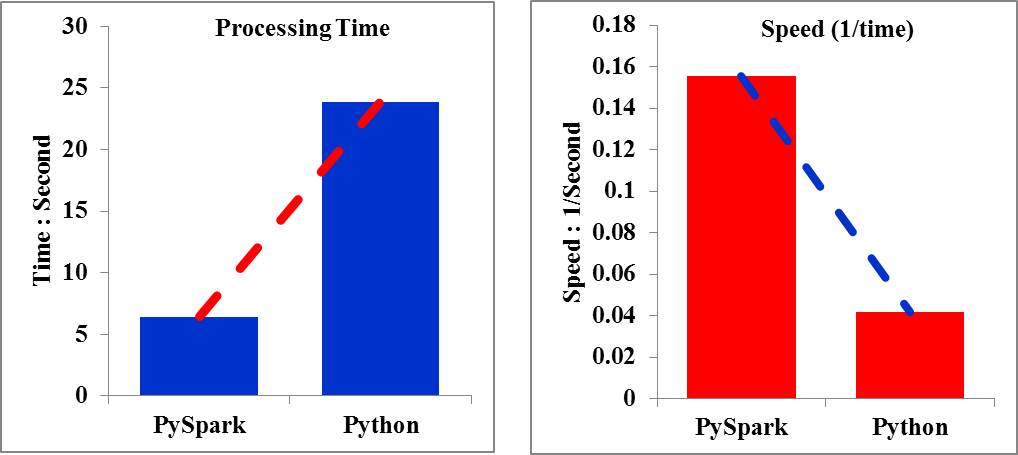}
	\caption{The processing time and speed comparison shows PySpark-based pipeline performs faster}
	\label{fig3}
\end{figure}

In this study we tested our proposed pipeline Fig.\ref{fig3} against the results from our previous papers \cite{sarraf2014brain}. Briefly, in the past studies, 7 males and 9 females with a mean age of 21.1 ± 2.2 years were recruited and structural and functional MRI data were collected. The standard fMRI preprocessing steps were applied to raw data using FMRIB Software Library v5.0 \cite{smith2004advances}. The goal of that study was to extract the brain networks (especially Default Mode Network) from independent components of brain imaging data. Probabilistic Independent Components of the preprocessed data were calculated by FSL-MELODIC resulted in 84 components. Next, using our template matching algorithm \cite{sarraf2014brain} the DMN was reconstructed from probabilistic independent components. In our brain network extractor and decision making algorithm, different methods such as normalized cross correlation, sum of squared differenced and dice coefficient. In our current study, we only used sum of squared error (SSD) in order to test our Spark solution.  In this work, we used PySpark Standalone version (http://spark.apache.org/) on single node to test and explore the potential application of our proposed pipeline. Those 84 brain components which were in Neuroimaging Informatics Technology Initiative (Nifti) format (standard format for NeuroImgaing data) were loaded into memory using Nibable package (http://nipy.org/nibabel) providing interfaces for neuroimaging data manipulation in Python. Next, the data in memory were converted to Resilient Distributed Datasets (RDD) format. RDD is a fundamental data structure of Spark. It is an immutable distributed collection of objects and ach dataset in RDD is divided into logical partitions, which may be computed on different nodes of the cluster. Formally, an RDD is a read-only, partitioned collection of records. RDDs can be created through deterministic operations on either data on stable storage or other RDDs. RDD is a fault-tolerant collection of elements that can be operated on in parallel. As mentioned above, we were to extract the default mode network from the components. Therefore, we used the DMN template developed by our team. The template was also loaded and converted to RDD.  SSD between 84 components and DMN templated were calculated using following equation (\ref{eq1}) in PySpark after flatmapping and zipping RDDs.

\begin{equation}\label{eq1}
SSD = \sum\limits_{x,y}[f(x,y)-t(x-u,y-v)]^2 
\end{equation} 
The performance was measured in PySpark which was equal to 6.43799 seconds. The same experiment was repeated in Python and the performance was measured and it was equal to 23.86625. This testing revealed that using PySpark big data platform even on single node runs the data faster than pure Python around 4 times (exactly 3.7) in our case. In addition, if the number of images increases the difference between PySpark and Python performance will increase as well. 
The Fig. 4 compares the measured performance between Python and PySpark which ran against our template matching script. Around 4 times speed up the running time is promising and we argue that our program was executed on a single node and was not completely designed for parallel computing and big data platform. In other words, if we develop our serial template matching algorithm for PySpark environment and parallel processing and also use high performance cluster instead of standalone - single node version of Spark, the performance will potentially improve up to 10 to 20 times especially when a huge dataset is processed.
\begin{table}[h]
	\caption{Comparison table between PySpark and Python in brain extraction application}
	\label{table_Comparison}
	\begin{center}
		\begin{tabular}{|c||c||c|}
			\hline
			Platform  & Time (second) & 1/Time\\
			\hline
			PySpark & 6.437999964 & 0.15533\\
			\hline
			Python & 23.86625409 & 0.0419\\
			\hline
		\end{tabular}
	\end{center}
\end{table}
\begin{figure}[h!]
	\includegraphics[width=\linewidth]{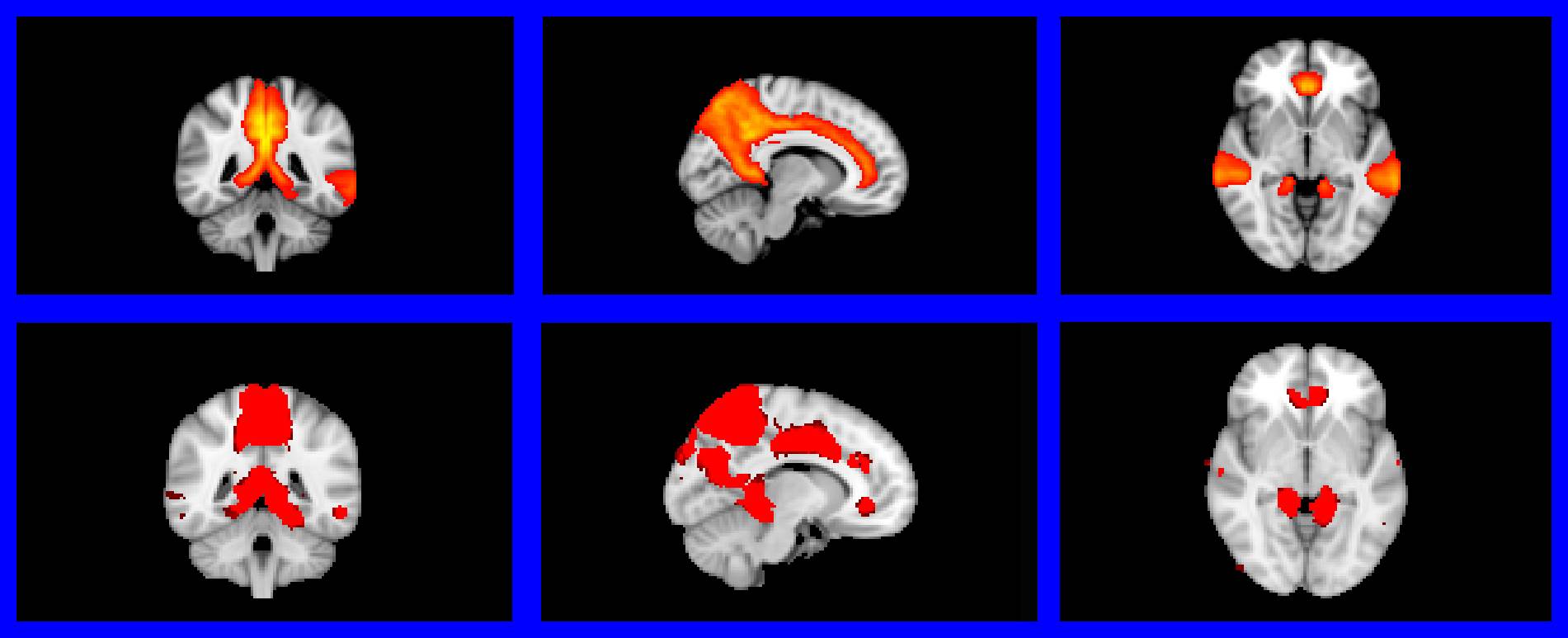}
	\caption{Using our algorithm described in [7] [9], we reconstructed the brain networks from independent components}
	\label{fig4}
\end{figure}
\begin{figure}[h!]
	\includegraphics[width=\linewidth]{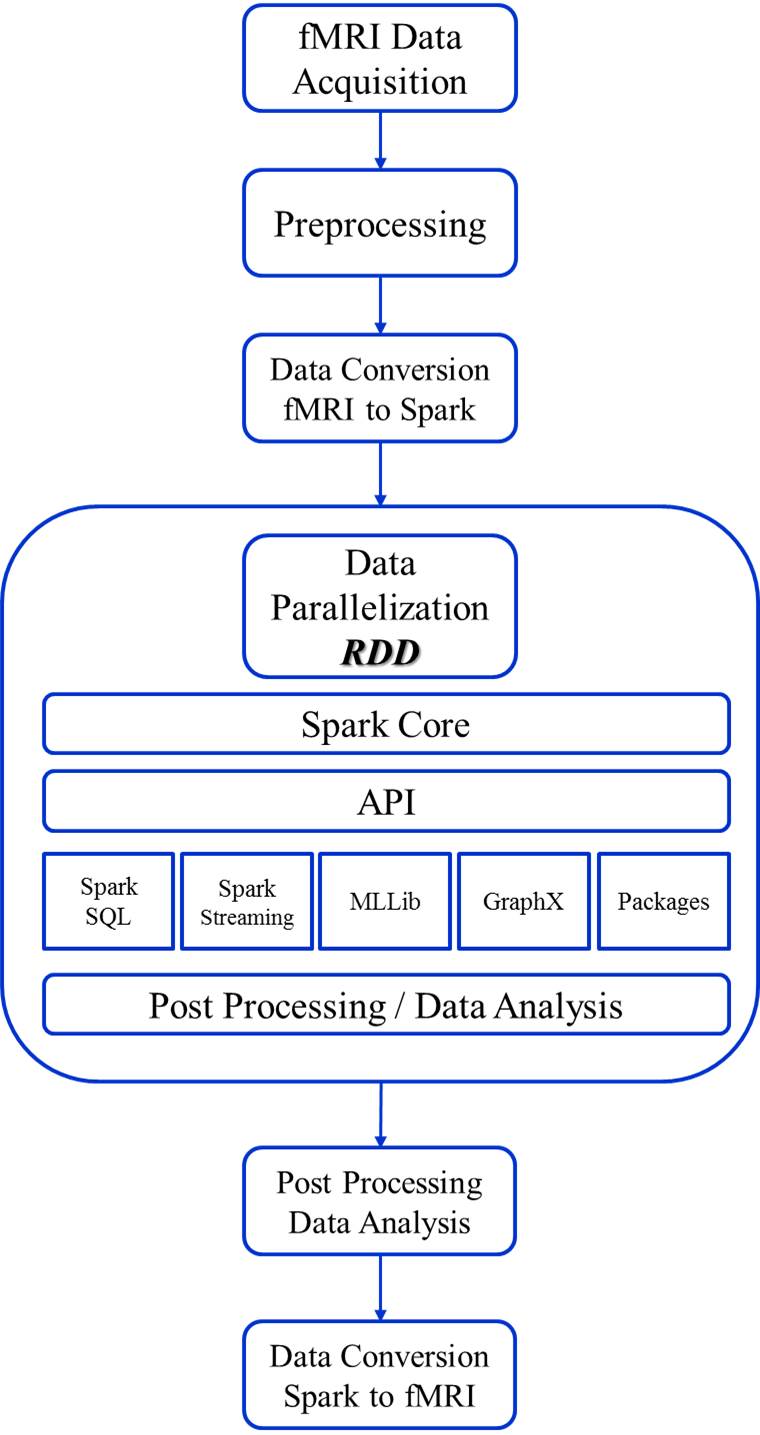}
	\caption{PySpark-based pipeline for fMRI data processing and analysis}
	\label{fig5}
\end{figure}

\section{CONCLUSION}
We developed and successfully tested our new PySpark-based pipeline on a single node to analyze functional MRI data for extracting brain networks. The new pipeline improved the processing time around 4 times faster than previous works while the accuracy remained at the same value. Furthermore, ease of use, in-memory data processing and storing results in different data structure are some important features of this pipeline. Also, this pipeline can easily expand to several nodes and high performance computing clusters for massive data analysis on large datasets which will definitely improve the processing time and the performance of the pipeline much more than a single node.
\section*{ACKNOWLEDGMENT} 
We would like to express our gratitude towards Drs. Ali Mohammad Golestani, Post-doctoral fellowship at Rotman Research Institute at Baycrest and Dr. Cristina Saverino Post-doctoral fellowship at Toronto Rehabilitation Institute-University Health Network for extending their help and support in this study.

\end{document}